\documentclass[prl,reprint,nofootinbib,superscriptaddress,preprintnumbers,showpacs]{revtex4-1}
\usepackage{amsfonts}
\usepackage{amssymb}
\usepackage{amsmath}
\usepackage{graphicx,color}
\usepackage{dcolumn}
\usepackage{epsfig}
\usepackage{bm}
\usepackage{bbm}
\usepackage{ulem}
\usepackage{hyperref}
\usepackage{lineno}
\usepackage{float}

\usepackage{color}

\def\gtap{\ \raise.3ex\hbox{$>$\kern-.75em\lower1ex\hbox{$\sim$}}\ }
\def\ltap{\ \raise.3ex\hbox{$<$\kern-.75em\lower1ex\hbox{$\sim$}}\ }

\allowdisplaybreaks[4]

\pdfoutput=1
\begin{document}

\title{
$X$ and $Z_{cs}$ in $B^+\to J/\psi\phi K^+$ as $s$-wave
threshold cusps and\\ alternative spin-parity assignments to $X(4274)$ and $X(4500)$
}
\author{Xuan Luo}
\affiliation{School of Physics and Optoelectronics Engineering, Anhui University, Hefei 230601, People's Republic of China}
\author{Satoshi X. Nakamura}
\email{satoshi@ustc.edu.cn}
\affiliation{
University of Science and Technology of China, Hefei 230026, 
People's Republic of China
}
\affiliation{
State Key Laboratory of Particle Detection and Electronics (IHEP-USTC), Hefei 230036, People's Republic of China}

\begin{abstract}
Recent LHCb's amplitude analysis on $B^+\to J/\psi \phi K^+$ 
suggests the existence of
exotic $X$ and $Z_{cs}$ hadrons, based on an assumption that 
Breit-Wigner resonances describe all the peak structures.
However, all the peaks and also dips in the spectra
are located at relevant meson-meson 
thresholds where 
threshold kinematical cusps might cause such structures.
This points to the importance of 
an independent amplitude analysis with due consideration of
the kinematical effects, and this is what we do in this work. 
Our model fits well
$J/\psi \phi$, $J/\psi K^+$, and 
$K^+\phi$ invariant mass distributions simultaneously,
demonstrating that 
all the $X$, $Z_{cs}$, and dip structures 
can be well described with the ordinary $s$-wave threshold cusps.
Spin-parity of the 
$X(4274)$ and $X(4500)$ structures are respectively $0^-$ and $1^-$ from our model, as opposed to 
$1^+$ and $0^+$ from the LHCb's.
With all relevant threshold cusps considered, the number of fitting
parameters seems to be significantly reduced.
The LHCb data requires 
$D_s^{(*)}\bar{D}^{*}$ scattering lengths 
in our model to be consistent with zero, 
disfavoring
$D_s^{(*)}\bar{D}^{*}$ molecule
interpretations of 
$Z_{cs}(4000)$ and $Z_{cs}(4220)$ and,
via the SU(3) relation,
being consistent with previous lattice QCD results.
\end{abstract}

\maketitle

{\it Introduction.}---
Recent experimental developments resulted in many discoveries of new
hadrons that are not categorized into the conventional $qqq$ and
$q\bar{q}$ structures. 
Countless theoretical papers followed to understand the nature of
such exotic hadrons often called $XYZ$, thereby 
deepening our knowledge of QCD in the nonperturbative regime;
see reviews~\cite{review_chen,review_olsen,review_Brambilla}.
Hadron properties such as mass, width, and spin-parity ($J^P$)
are crucial information to address the hadrons' nature and structures,
and amplitude analysis is the method to extract those information from
data. 
However, 
amplitude analysis results are often neither unique nor model-independent for 
assumptions and simplifications that go into the
analyses.
It is therefore
important to bring different and independent analysis results together
to establish the hadron properties through critical reviews and comparisons.

The $B^+\to J/\psi \phi K^+$ decay~\footnote{We follow the hadron naming scheme in Ref.~\cite{pdg}. 
We often denote 
$J/\psi$ and $\psi(2S)$ 
by $\psi$ and $\psi'$, respectively, for simplicity.
$D_{s0}^*(2317)$ and $D_{s1}(2536)$ are generically denoted
by $D_{sJ}^{(*)}$.
The charge conjugate decays are implied throughout, and 
charge indices are often suppressed.
} is an interesting case. 
Earlier analyses~\cite{cdf,belle,cdf2,lhcb_old,cms,d0,babar,d02} fitted structures in 
the $J/\psi\phi$ invariant mass
($M_{J/\psi\phi}$) distribution with Breit-Wigner amplitudes, and claimed exotic 
$X(4140)$ and $X(4274)$ without $J^P$ determinations.
%
A first six-dimensional amplitude analysis was done by 
the LHCb Collaboration~\cite{lhcb_phi1,lhcb_phi2},
and four $X$ states with $J^P$ were reported:
$X(4140)$ and $X(4274)$ with $J^P=1^+$; 
$X(4500)$ and $X(4700)$ with $J^P=0^+$.
These $X$ states were confirmed with 
higher statistics data recently, 
and
$1^+ X(4685)$,
$2^- X(4150)$, and 
$1^- X(4630)$ were also added~\cite{lhcb_phi}.
Furthermore, the LHCb claimed 
$1^+ cu\bar{c}\bar{s}$ tetraquarks
$Z_{cs}(4000)^+$ and $Z_{cs}(4220)^+$
appearing as bumps in the
$M_{J/\psi K^+}$ distribution.

The LHCb's analysis assumes that all bumps in 
the $M_{J/\psi\phi}$ and 
$M_{J/\psi K^+}$ distributions are caused by 
$X$ and $Z_{cs}$ resonances that can be
simulated by Breit-Wigner amplitudes. 
However, these $X$ [$Z_{cs}$] bumps and also dips are located at
$D_s^*\bar{D}_s^{(*)}$,
$D_{sJ}^{(*)}\bar{D}_s^{(*)}$, 
and $\psi'\phi$ 
[$D_{s}^{(*)}\bar{D}^*$]
thresholds
where kinematical effects such as threshold cusps and/or triangle singularities may
cause resonancelike and dip structures~\cite{ts_review}.
Indeed,
it has been shown that $X(4140)$ and $X(4700)$ can be described with 
$D^*_{s}\bar{D}_s$ and $\psi'\phi$
threshold cusps, respectively~\cite{lhcb_phi2,swanson,xhliu,xkdong2,ortega1,Nakamura:2021bvs}.
While $1^+ X(4274)$ [$0^+ X(4500)$] at
the $D_{s0}^{*}(2317)\bar{D}_s$
[$D_{s1}(2536)\bar{D}_s$] threshold
cannot be an ordinary $s$-wave cusp for having different $J^P$,
they might still be described
with $p$-wave cusps 
enhanced by quasi double-triangle singularities~\cite{Nakamura:2021bvs}.
It is however noted that
the LHCb's $J^P$ assignments are not model-independent
but influenced by their assumptions.
Once possible threshold cusps not only at the peaks but also
at the dips are considered in the fit, 
it is unclear whether 
the LHCb's $J^P$ assignments remain unchanged.
We address this issue.

\begin{figure*}[t]
\begin{center}
\includegraphics[width=0.25\textwidth]{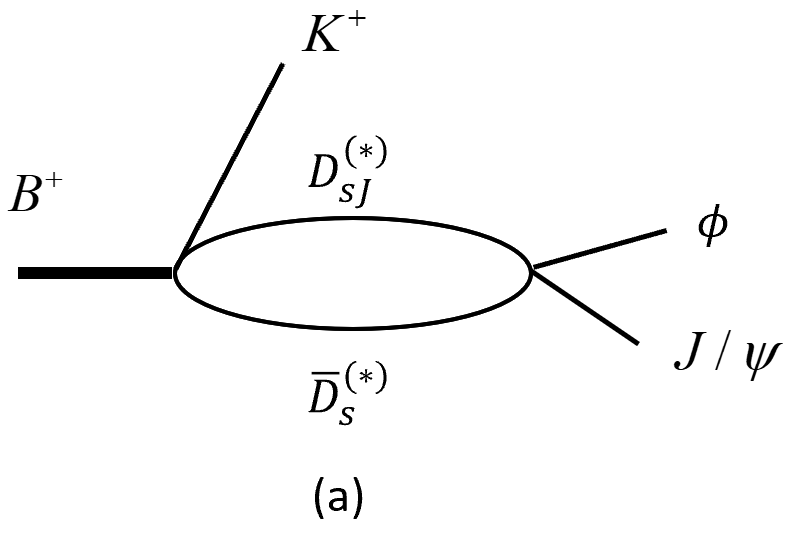}
\includegraphics[width=0.25\textwidth]{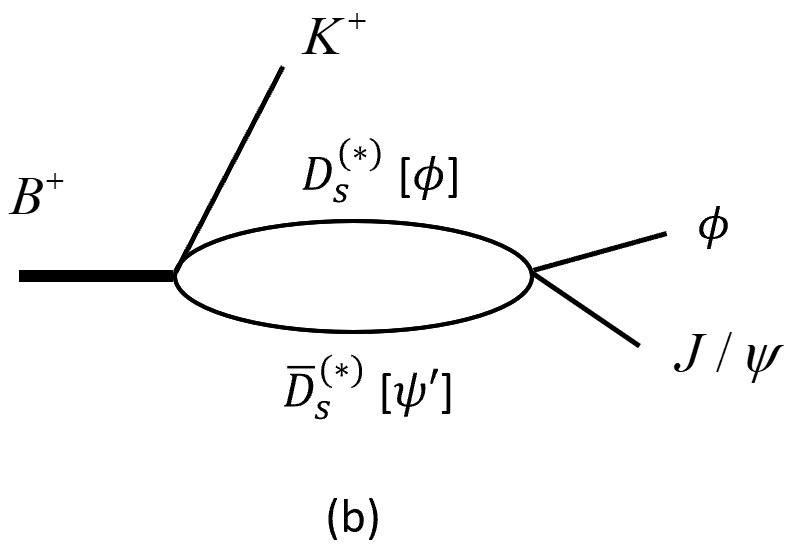}
\includegraphics[width=0.25\textwidth]{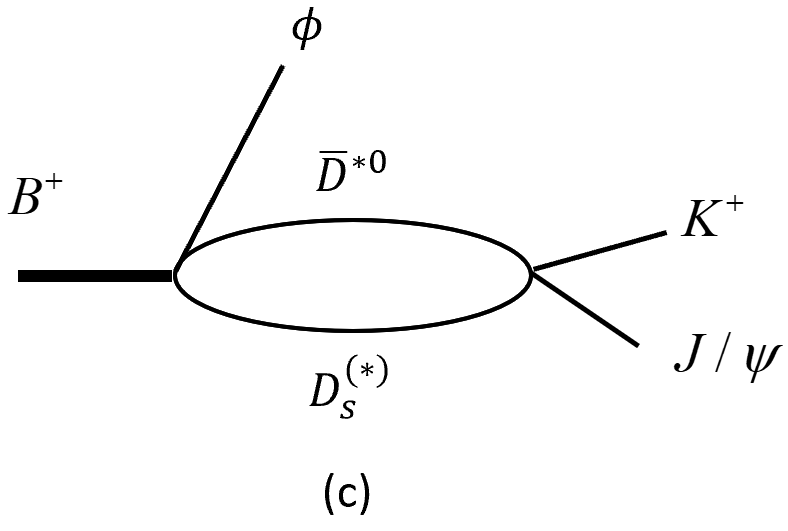}
\includegraphics[width=0.20\textwidth]{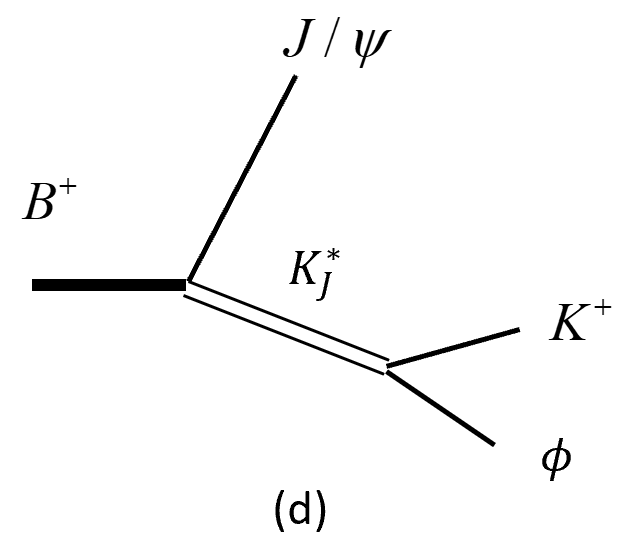}
\end{center}
 \caption{
$B^+\to J/\psi \phi K^+$ 
mechanisms:
(a) $D_{sJ}^{(*)+}{D}_s^{(*)-}$ ($0^-, 1^-$) one-loop;
(b) $D_{s}^{*+}{D}_s^{(*)-}$ and $\psi'\phi$ ($0^+, 1^+$) one-loop;
(c) $D_{s}^{(*)+}\bar{D}^{*0}$ ($1^+$) one-loop ;
(d) $K^*_J$ ($K$, $K^*, K_1$, $K_2$) excitations.
 }
\label{fig:diag}
\end{figure*}

Another issue concerns the nature of the $Z_{cs}(4000)$ and $Z_{cs}(4220)$.
A similar structure, called $Z_{cs}(3985)$,
was also discovered by
the BESIII collaboration 
in $e^+ e^- \to K^+(D_s^- D^{*0}+D_s^{*-} D^0)$~\cite{BESIII:2020qkh}.
While $Z_{cs}(4000)$ and $Z_{cs}(3985)$ have similar masses
($4003\pm 6^{+4}_{-14}$~MeV and $3982.5^{+1.8}_{-2.6}\pm 2.1$~MeV),
their widths are rather different
($131\pm 15\pm 26$~MeV and $12.8^{+5.3}_{-4.4}\pm 3.0$~MeV);
the first (second) errors are statistical (systematic).
$Z_{cs}(3985)$ and 
$Z_{cs}(4000)$ are argued to be the same $cu\bar{c}\bar{s}$ 
tetraquark state in Refs.~\cite{pshi21,Giron:2021sla}.
However, other works 
considered them to be different tetraquark states~\cite{Wang:2020rcx,maiani21,jbwang22,rosner21},
or different $D_s^{(*)}\bar{D}^{(*)}$ molecules~\cite{shan22,lmeng21,Wang:2020rcx},
or one of them is a tetraquark and the other is a molecule~\cite{zgwang22}.
$Z_{cs}(3985)$ and $Z_{cs}(4000)$ may also be from a common 
virtual pole that enhances 
the $D_s\bar{D}^{*}$ threshold cusp~\cite{zcs_model1,zyang21},
as demonstrated by fitting
both the LHCb's $M_{J/\psi K^+}$ distribution and BESIII data~\cite{zcs_model1}.
Also, $J/\psi K^{*+}$ 
and $\psi^\prime K^+$ threshold cusps could cause
the $Z_{cs}(4000)$ and $Z_{cs}(4220)$ structures, respectively~\cite{Ge:2021sdq}.

$Z_{cs}(3985/4000)$ may be regarded as a SU(3) partner of 
$Z_{c}(3900)$~\cite{lmeng21,zcs_model1,maiani21,pshi21,mdu22,vbaru22}.
Lattice QCD (LQCD) results disfavor the existence of a 
narrow $Z_{c}(3900)$ pole,
suggesting $Z_{c}(3900)$ to be a kinematical effect~\cite{Prelovsek13,ychen14,Prelovsek15,ikeda16,cheung17}.
This implies, via the SU(3) relation, no pole for 
$Z_{cs}(3985)$ and/or $Z_{cs}(4000)$.
However, consistency with the LQCD results was not
considered in most previous models~\footnote{
An exception is
Refs.~\cite{Albaladejo1,Albaladejo2}.
}.

In this work, we develop a model that simultaneously describes 
the ${J/\psi\phi}$, ${J/\psi K^+}$, and ${K^+\phi}$ invariant mass
distributions for $B^+ \to J/\psi \phi K^+$ from the LHCb.
We demonstrate that 
all the peaks ($X$, $Z_{cs}$) and dips in the 
$M_{J/\psi\phi}$ and $M_{J/\psi K^+}$ distributions 
are well described
with ordinary $s$-wave threshold cusps from one-loop diagrams in Fig.~\ref{fig:diag};
virtual poles near the thresholds are not necessary
for a good fit. 
Our model, $J^P$ of the cusps as well, should be well-constrained by 
simultaneously fitting the three invariant mass distributions.
Thus, we claim
$J^P=0^-$ and $1^-$ for the $X(4274)$ and $X(4500)$ cusps, respectively,
alternative to $J^P=1^+$ and $0^+$ from the LHCb analysis;
the different $J^P$ assignments would be from considering different mechanisms.
We will argue possible advantages of our model over the LHCb's model. 
We also examine to what extent 
the $D_s^{(*)}\bar{D}^{*}$ molecule interpretation of 
$Z_{cs}(4000)$ and $Z_{cs}(4220)$
is allowed by the LHCb data. 
The $D_s^{(*)}\bar{D}^{*}$ scattering lengths in our model is required to be consistent with
zero, disfavoring the molecule interpretation, and being consistent with
the above-mentioned LQCD results.

{\it The model.}---
We consider 
one-loop mechanisms of 
Fig.~\ref{fig:diag}(a,b)
[Fig.~\ref{fig:diag}(c)]
and their $s$-wave threshold cusps
that generate structures in the $M_{J/\psi\phi}$ [$M_{J/\psi K^+}$]
distribution of $B^+\to J/\psi \phi K^+$.
We also consider $K^*_J$ excitation mechanisms of Fig.~\ref{fig:diag}(d)
that would shape the $M_{K^+\phi}$ distribution.
We assume that other possible mechanisms play a minor role, and their
effects can be effectively absorbed by the considered mechanisms. 
We derive the corresponding amplitudes
by writing down effective Lagrangians of relevant hadrons and their 
matrix elements,
and combining them following the time-ordered perturbation theory.

The one-loop mechanisms of Fig.~\ref{fig:diag}(a)
include $s$-wave pairs of
\begin{eqnarray}
D_{s0}^*(2317)^+{D}^-_s(0^-), \ \
D_{s0}^*(2317)^+{D}^{*-}_s(1^-), \nonumber\\
D_{s1}(2536)^+{D}^-_s(1^-), \ \
D_{s1}(2536)^+{D}^{*-}_s(0^-),
\label{eq:ddpair}
\end{eqnarray}
where $J^P$ of a pair is indicated
in the parenthesis;
a $J^P=0^-$ ($1^-$) pair is from a parity-violating (conserving) weak
decay. 
These mechanisms include
short-range (e.g., quark-exchange) 
$D_{sJ}^{(*)}\bar{D}_s^{(*)}\to J/\psi\phi$ interactions
that would require a $c\bar{s}$ component in $D_{sJ}^{(*)}$.
$D_{s1}(2536)$ is considered to be a $p$-wave
$c\bar{s}$~\cite{Yang_Ds0}.
While $D_{s0}^*(2317)$ may have 
a dominant $DK$-molecule component as found by 
analyzing LQCD energy spectrum~\cite{liu_Ds0,torres_Ds0,Cheung_Ds0,Yang_Ds0},
a bare $c\bar{s}$ component can still be an important
constituent~\cite{Yang_Ds0}.
The diagrams of Fig.~\ref{fig:diag}(b)
include $s$-wave pairs of
\begin{eqnarray}
&&D_{s}^{*+}{D}^-_s(1^+), \ 
D_{s}^{*+}{D}^{*-}_s(0^+), \ 
\psi'\phi(0^+), \ 
\psi'\phi(1^+),
\label{eq:ddpair2}
\end{eqnarray}
where a $J^P=1^+$ ($0^+$) pair is for a parity-violating (conserving) process.
Since $D_s^*\bar{D}^*_s(1^+)$ and 
$J/\psi\phi(1^+)$ have different 
$C$-parity,
$D_s^*\bar{D}^*_s(1^+)$ does not contribute here. 
The diagrams of Fig.~\ref{fig:diag}(c)
include $s$-wave pairs of
\begin{eqnarray}
D_s^+\bar{D}^{*0}(1^+), \ \ 
D_s^{*+}\bar{D}^{*0}(1^+),
\label{eq:ddpair3}
\end{eqnarray}
that can contribute to both 
parity-conserving and violating processes.
While a $D_s^{*+}\bar{D}^{0}(1^+)$ one-loop mechanism 
is also possible, its singular behavior is similar to that of 
$D_s^+\bar{D}^{*0}(1^+)$ due to almost degenerate thresholds 
($\sim$1.8~MeV difference). 
We thus assume that the 
$D_s^+\bar{D}^{*0}(1^+)$ one-loop amplitude implicitly absorbs the 
$D_s^{*+}\bar{D}^{0}(1^+)$ contribution.

In Eqs.~(\ref{eq:ddpair})-(\ref{eq:ddpair3}), we did not exhaust all
possible $J^P$ such as 
$D_{s1}(2536)^+{D}^{*-}_s(1^-, 2^-)$ and $D_{s}^{*+}{D}^{*-}_s(2^+)$.
While they can in principle contribute to the process, we found them
unnecessary to reasonably fit the three invariant mass distributions. 
We thus do not consider them and keep the number of fitting parameters smaller.
Also, we do not explicitly consider 
charge analogous amplitudes that include, for example,  
${D}^+_s D_{s0}^*(2317)^-$ rather than 
$D_{s0}^*(2317)^+{D}^-_s$ in Fig.~\ref{fig:diag}(a).
While the charge analogous amplitudes
generally have independent strengths,
their singular behaviors are the same as the original ones.
It is understood that 
their effects and projections onto positive $C$-parity
are taken into account in
coupling strengths of the considered processes.

\begin{table}[t]
\renewcommand{\arraystretch}{1.4}
\tabcolsep=1.5mm
\caption{\label{tab:KJ} $K^*_J$ in Fig.~\ref{fig:diag}(d).
The first row indicates $J^P$ of $K^*_J$.
In the default model, parity-conserving (pc) and/or 
-violating (pv) amplitudes 
or neither ($-$) are considered, as indicated in the 
square brackets.
}
\begin{tabular}{cccc}\hline
 $0^-$& $1^-$& $1^+$& $2^-$  \\ 
$K(1460)[-]$&  $K^*(1410)[\rm pc]$&  $K_1(1400)[-]$ & $K_2(1770)[\rm pc]$ \\
            &  $K^*(1680)[\rm pc]$&  $K_1(1650)[\rm pc, pv]$ &
	     $K_2(1820)[\rm pc]$ \\\hline
\end{tabular}
\end{table}

We consider the $K^*_J$-excitation mechanisms of 
Fig.~\ref{fig:diag}(d) in Breit-Wigner forms.
With the LHCb's amplitude analysis result as reference, 
we consider 
$K^*_J$ as listed in Table~\ref{tab:KJ}.
Each $K^*_J$ may have
parity-conserving and/or -violating $B^+ \to K^*_J J/\psi$ couplings,
depending on $J^P$ of $K^*_J$.

We present an amplitude formula for
Fig.~\ref{fig:diag}(c)
with a $D_{s}^+ \bar{D}^{*0}(1^+)$ pair
that generates a $Z_{cs}(4000)$-like cusp;
see the Supplemental Material for amplitude formulas for other mechanisms.
We denote the energy, width, three-momentum and 
polarization vector of a particle $x$ by
 $E_x$, $\Gamma_x$, $\vec p_x$ and $\vec \varepsilon_x$, respectively.
The particle masses and widths are taken from 
Ref.~\cite{pdg} unless otherwise stated. 
A parity-conserving (pc) $B^+ \to D_s^+ \bar{D}^{*0}\phi$ vertex
and the subsequent $D_s^+\bar{D}^{*0}\to J/\psi K^+$ interaction
 that enter the amplitude are
\begin{align}
&c_{D_s\bar{D}^{*0} (1^+)}^{\rm pc}\vec \varepsilon_{\bar{D}^{*0}} \cdot \vec \varepsilon_\phi F^{00}_{D_s\bar{D}^{*0} \phi,B},
\\
&c_{\psi K, D_s\bar{D}^{*0}}^{1^+} \vec \varepsilon_{\bar{D}^{*0}}
 \cdot \vec \varepsilon_{\psi} f_{\psi K}^0 f^0_{D_s\bar{D}^{*0}},
\end{align}
respectively,
where we introduced dipole form factors 
$F_{ijk,l}^{LL'}$ and $f_{ij}^{L}$;
we use a common cutoff of $\Lambda=1$~GeV in all form factors;
$c_{D_s\bar{D}^{*0} (1^+)}^{\rm pc}$ and 
$c_{\psi K, D_s\bar{D}^{*0}}^{1^+}$
are coupling constants.
With the above ingredients, the one-loop amplitude is given by
\begin{eqnarray}
\label{eq:zcs4000amp}
A_{\bar{D}^{*0}D_s(1^+)}^{\rm 1L,pc} &=&
c_{\psi K, D_s\bar{D}^{*0}}^{1^+}
c_{D_s\bar{D}^{*0} (1^+)}^{\rm pc}
\vec \varepsilon_\psi \cdot \vec \varepsilon_\phi 
\nonumber \\
&&\times \int d^3 p_{D_s} {
f_{\psi K}^0 f^0_{D_s\bar{D}^{*0}}F^{00}_{D_s\bar{D}^{*0}
\phi,B}\over 
M_{\psi K}-E_{D_s}-E_{\bar{D}^{*0}}+i\varepsilon},
\end{eqnarray}
where $\Gamma_{D^{*0}}$ has been neglected for being estimated to
be small ($\Gamma_{D^{*0}}\sim 55$~keV~\cite{sxn_x}).

The $D_s^{(*)+} \bar{D}^{*0}$ threshold cusps
from Eq.~(\ref{eq:zcs4000amp})
 could be enhanced by
virtual or bound states near the thresholds~\cite{xkdong}.
To implement this effect,
we describe the $D_s^{(*)+} \bar{D}^{*0}\to J/\psi K^+$ transition
with a single-channel $D_s^{(*)+} \bar{D}^{*0}$ scattering
followed by a perturbative 
$D_s^{(*)+} \bar{D}^{*0}\to J/\psi K^+$ transition.
We use a $D_s^{(*)+} \bar{D}^{*0}$ interaction potential of
\begin{eqnarray}
v_\alpha (p',p) &=& f^0_\alpha(p') h_\alpha\; f^0_\alpha(p) ,
\label{eq:cont-ptl}
\end{eqnarray}
where $\alpha$ labels an interaction channel;
$h_\alpha$ is a coupling constant and $f^L_\alpha$ is a dipole 
form factor.
We can implement the rescattering effect 
in Eq.~(\ref{eq:zcs4000amp})
by multiplying 
$[1 - h_\alpha \sigma_\alpha(M_{J/\psi K^+})]^{-1}$
 with
\begin{eqnarray}
\sigma_\alpha(E) &=&
 \int\! dq q^2 
{ \left[f^0_\alpha(q)\right]^2 
\over E-E_{D_s}(q)-E_{\bar{D}^{*0}}(q) 
+ i\varepsilon} .
\label{eq:sigma}
\end{eqnarray}
The default model
does not include the rescattering effects ($h_\alpha=0$)
for Fig.~\ref{fig:diag}(a,c).
We will examine the rescattering effect on the $Z_{cs}$ structures separately. 

Meanwhile, our default model
includes similar rescattering effects in
the $D_s^{*} \bar{D}_s^{(*)}\to J/\psi \phi$ transitions
of Fig.~\ref{fig:diag}(b).
The $D_s^{*} \bar{D}_s^{(*)}$ interaction strengths 
are chosen to be moderately attractive ($h_\alpha=-2$).
The scattering length is
$a\sim 0.55$~fm~\footnote{
The scattering length $(a)$ is related to the phase shift $(\delta)$ by 
$p\cot\delta=1/a + {\cal O}(p^2)$.}, 
and a virtual pole is located at $\sim 20$~MeV below the 
$D_s^{*} \bar{D}_s^{(*)}$ threshold.
In Ref.~\cite{xkdong2}, the authors used 
a contact $D_s^{*} \bar{D}_s^{(*)}$
interaction saturated by a $\phi$-exchange mechanism, and 
found similar virtual poles.

\begin{figure}[b]
\begin{center}
\includegraphics[width=0.225\textwidth]{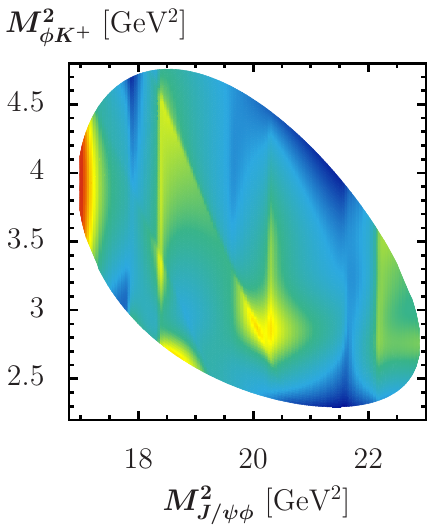}
\includegraphics[width=0.252\textwidth]{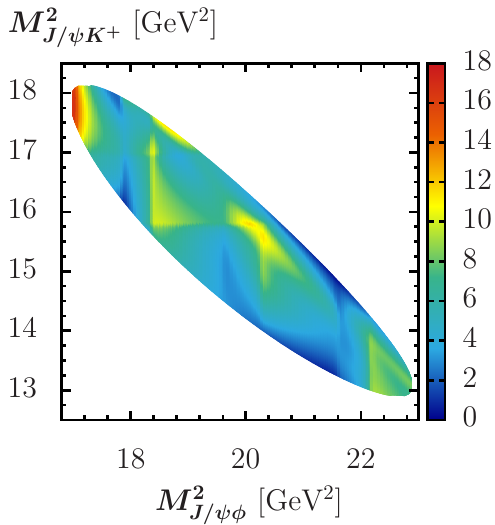}
\end{center}
\caption{$B^+ \to J/\psi \phi K^+$ Dalitz plot distributions from the default model.
No smearing is applied.
}
\label{fig:dalitz}
\end{figure}

\begin{figure*}[htbp]
\begin{center}
\includegraphics[width=0.495\textwidth]{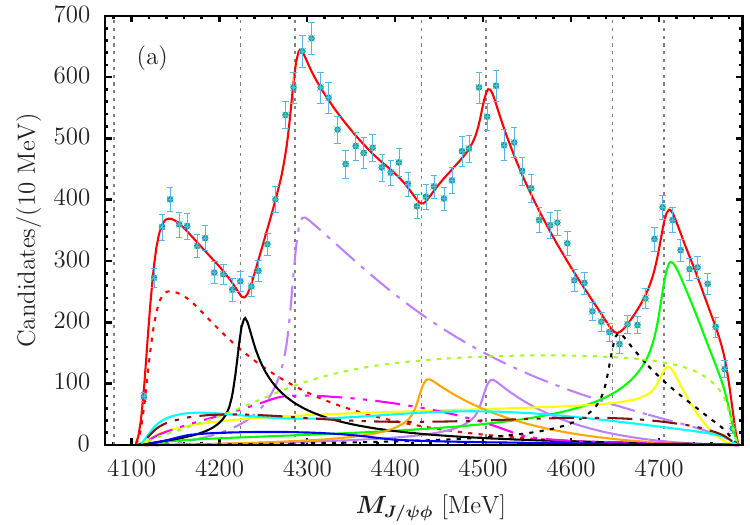}
\includegraphics[width=0.495\textwidth]{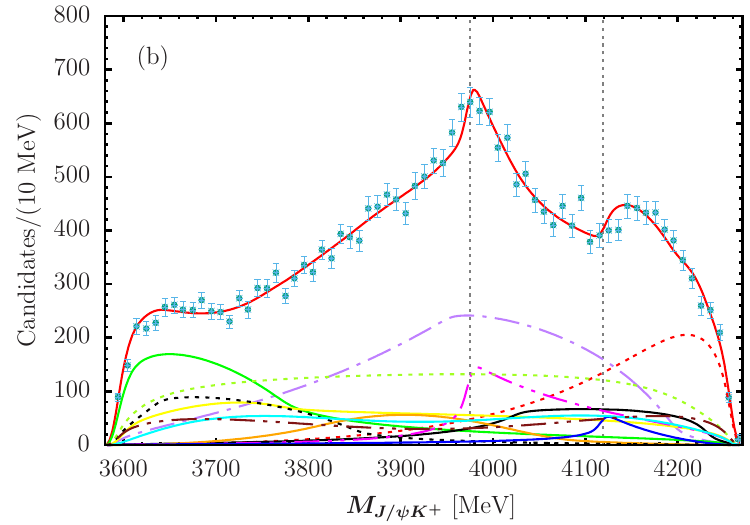}
\end{center}
\caption{\label{fig:res}
Combined fit to
(a) $J/\psi \phi$, (b) $J/\psi K^+$, and 
(c) $\phi K^+$ invariant mass distributions for 
$B^+ \to J/\psi \phi K^+$.
The red solid curves are from the default model.
Contributions are from Fig.~\ref{fig:diag}(a-c) that include
$D_{s}^{*+}{D}^-_s(1^+)$ [red dashed],
$D_{s}^{*+}{D}^{*-}_s(0^+)$ [black solid],
$D_{s0}^*(2317)^+{D}^-_s(0^-)$ [purple dash-dotted],
$D_{s0}^*(2317)^+{D}^{*-}_s(1^-)$ [orange solid],
$D_{s1}(2536)^+{D}^-_s(1^-)$ [purple solid],
$D_{s1}(2536)^+{D}^{*-}_s(0^-)$ [black dashed],
$\psi'\phi(0^+)$ [green solid],
$\psi'\phi(1^+)$ [yellow solid],
$D_s^+\bar{D}^{*0}(1^+)$ [magenta dash-two-dotted], and
$D_s^{*+}\bar{D}^{*0}(1^+)$ [blue solid].
Contributions from Fig.~\ref{fig:diag}(d) 
are $K^*$ [cyan solid], $K_1$ [green dashed], and $K_2$ [brown
dash-two-dotted].
The dotted vertical lines in (a) [(b)]
indicate thresholds for,
from left to right, 
$D_{s}^*\bar{D}_s$,
$D_{s}^*\bar{D}^*_s$,
$D_{s0}^*(2317)\bar{D}_s$,
$D_{s0}^*(2317)\bar{D}^*_s$,
$D_{s1}(2536)\bar{D}_s$,
$D_{s1}(2536)\bar{D}^*_s$, and
$\psi'\phi$
[$D_s^+\bar{D}^{*0}$ and $D_s^{*+}\bar{D}^{*0}$],
respectively.
The data are from Ref. \cite{lhcb_phi}.
}
\end{figure*} 

\begin{figure}[htbp]
\begin{center}
\includegraphics[width=0.50\textwidth]{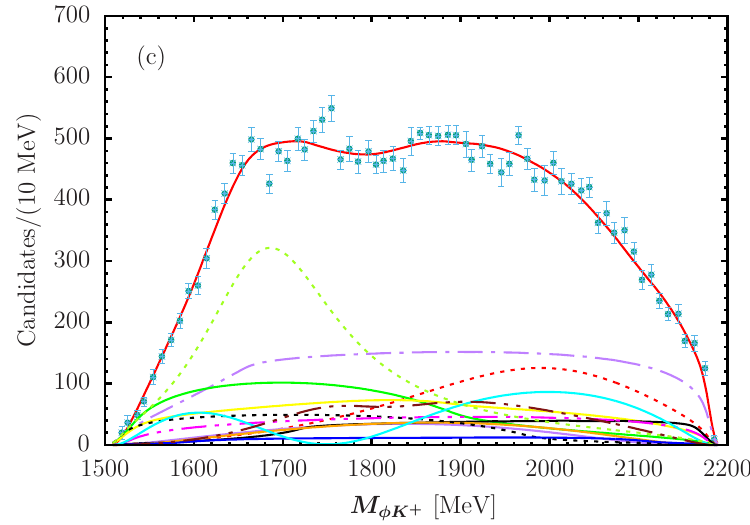}
\caption{\label{fig:res2}
Continued from Fig.~\ref{fig:res}.
}
\end{center}
\end{figure}

{\it Results and discussions.}---
We simultaneously fit 
the $M_{J/\psi\phi}$, $M_{J/\psi K^+}$, 
and $M_{K^+\phi}$ distributions
of $B^+\to J/\psi \phi K^+$ from the LHCb 
using the model described above.
As seen in Eq.~(\ref{eq:zcs4000amp}),
each amplitude has 
a complex overall factor from the product of coupling
constants. 
We determine the complex factors by fitting the data
since other experimental inputs are lacking.
During the fit, we remove relatively unimportant mechanisms
to reduce the fitting parameters and retain essential mechanisms.
For Fig.~\ref{fig:diag}(c) with 
Eq.~(\ref{eq:ddpair3}), we remove a parity-violating [conserving] one with
$D_s^+\bar{D}^{*0}(1^+)$ [$D_s^{*+}\bar{D}^{*0}(1^+)$].
Among the $K^*_J$-excitation mechanisms, 
we retain only those indicated by 'pc' and/or 'pv' in Table~\ref{tab:KJ}.
Our default model totally has 16 mechanisms, and 
$2\times 16 -3 = 29$ fitting parameters where $-3$ is from the
arbitrariness of 
the absolute normalization of the full amplitude, and that of 
overall phases of the parity-conserving and -violating full amplitudes.
The parameter values and fit fractions for the default
model are provided in the Supplemental Material.

We first present Dalitz plot distributions from the default
model in Fig.~\ref{fig:dalitz}.
Comparing with the LHCb's Dalitz plots~\cite{lhcb_phi},
the overall patterns are similar.
Since our plots are not smeared with the experimental resolution, 
the peak structures seem sharper than the data.

In Figs.~\ref{fig:res} and \ref{fig:res2},
our default model (red solid curves) is shown to 
agree well with the LHCb data for the $M_{J/\psi\phi}$, $M_{J/\psi K^+}$, 
and $M_{K^+\phi}$ distributions;
$\chi^2/{\rm ndf}=(102.3+94.2+113.7)/(3\times 68-29)=1.77$ 
where three $\chi^2$ values are from 
the $M_{J/\psi\phi}$, $M_{J/\psi K^+}$, 
and $M_{K^+\phi}$ distributions, respectively, and
ndf is the number of the
bins subtracted by the number of fitting parameters.
All theoretical curves are smeared with the experimental bin width.
The $X(4140)$, $X(4274)$, $X(4500)$, $X(4700/4685)$ peaks
in the $M_{J/\psi\phi}$ distribution
are well described by the 
$D_{s}^{*+}{D}^-_s(1^+)$ [red dashed curve],
$D_{s0}^*(2317)^+{D}^-_s(0^-)$ [purple dash-dotted], 
$D_{s1}(2536)^+{D}^-_s(1^-)$ [purple solid], and 
$\psi'\phi(0^+/1^+)$ [green solid/yellow solid]
threshold cusps, respectively.
Also, the three dips are well fitted with the
$D_{s}^{*+}{D}^{*-}_s(0^+)$ [black solid],
$D_{s0}^*(2317)^+{D}^{*-}_s(1^-)$ [orange solid], and
$D_{s1}(2536)^+{D}^{*-}_s(0^-)$ [black dashed]
threshold cusps.
The cusp peak positions are slightly above the thresholds
due to smearing the asymmetric cusp shapes.

In the $M_{J/\psi K^+}$ distribution, 
the $D_s^+\bar{D}^{*0}(1^+)$ threshold
cusp [magenta dash-two-dotted]
fits well the $Z_{cs}(4000)$-like peak.
The $D_s^{*+}\bar{D}^{*0}(1^+)$ threshold
cusp [blue solid]
creates a dip at 
$M_{J/\psi K^+}\sim 4120$~MeV and, 
combined with the shrinking phase-space near the kinematical endpoint, 
the $Z_{cs}(4220)$-like structure is formed.
While the $K^*_J$-excitation mechanisms do not create noticeable structures in 
the $M_{K^+\phi}$ distribution,
their contributions and interferences are important for a reasonable
fit. 

\begin{figure}[htbp]
\begin{center}
\includegraphics[width=0.5\textwidth]{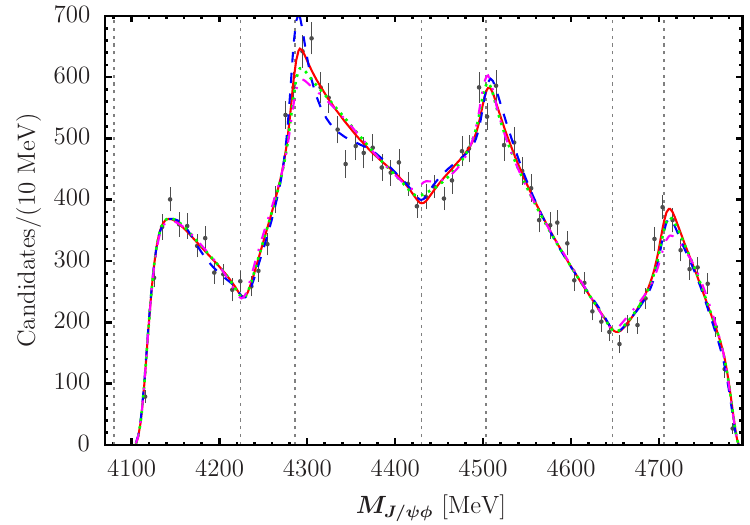}
\caption{\label{fig:ff}
The $J/\psi \phi$ invariant mass distributions from the fits
with different cutoff ($\Lambda$) values
in the dipole form factors.
The blue dashed, red solid, green dotted, and magenta dash-dotted curves are obtained
with $\Lambda=750, 1000, 1250$, and 1500~MeV, respectively.
Other features are the same as those in Fig.~\ref{fig:res}(a).
}
\end{center}
\end{figure}

\begin{figure}[htbp]
\begin{center}
\includegraphics[width=0.5\textwidth]{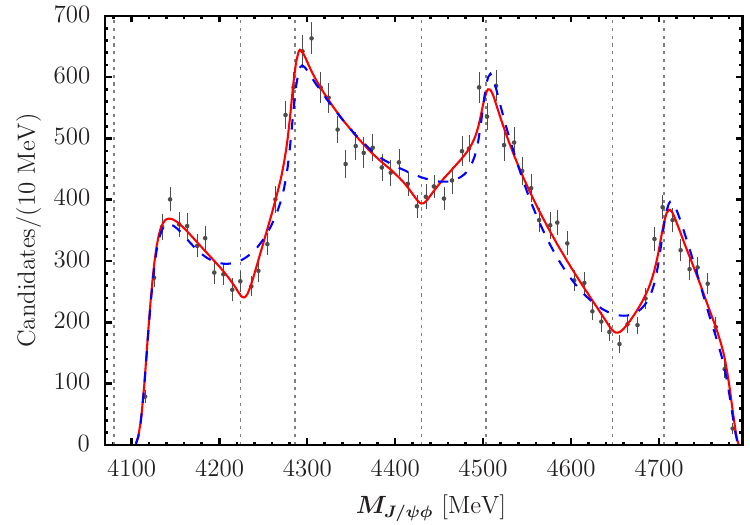}
\caption{\label{fig:dip}
The $J/\psi \phi$ invariant mass distributions 
from the fits with different mechanisms.
The red solid curve is from the default model.
The blue dashed curve is from a model where 
the $D_{s}^{*+}{D}^{*-}_s(0^+)$, 
$D_{s0}^*(2317)^+{D}^{*-}_s(1^-)$, and
$D_{s1}(2536)^+{D}^{*-}_s(0^-)$ loop mechanisms [Fig.~\ref{fig:diag}(a,b)]
are removed from the default mechanisms.
Other features are the same as those in Fig.~\ref{fig:res}(a).
}
\end{center}
\end{figure}

We examine if the fit is stable against changing the form
factor.
Instead of $\Lambda=1000$~MeV (cutoff) in all the dipole form factors 
of the default model, 
we fit the data with 
$\Lambda=750$, 1250, and 1500~MeV.
As seen in Fig.~\ref{fig:ff}
for the $M_{J/\psi\phi}$ distribution, 
while the sharpness of the $X(4274)$ peak is somewhat sensitive to the cutoff value,
the fit is reasonably stable overall.
Similarly, stable fits are also obtained 
for the $M_{J/\psi K^+}$ and $M_{K^+\phi}$ distributions.
This stability is expected since the threshold cusps are caused by 
low-momentum components in the loop integrals, and are insensitive to
how high-momentum components are cut off.
We also used monopole and Gaussian form factors 
with $\Lambda=1$~GeV, and 
confirmed that the result is very similar to the case of 
$\Lambda=1250$~MeV in Fig.~\ref{fig:ff}.

Our results are different from the LHCb's in many points.
First, all $X$ and $Z_{cs}$ structures are from the threshold cusps in
our model, while they are from resonances of the Breit-Wigner forms in
the LHCb's.
Second, $J^P$ of the 
$X(4274)$ and $X(4500)$ peaks are respectively
$0^-$ and $1^-$ cusps in our model while 
$1^+$ and $0^+$ resonances in the LHCb's.
This difference in $J^P$ might be from the fact that 
our model creates the sharp three dips
in the $M_{J/\psi\phi}$ distribution
with the threshold cusps.
In Fig.~\ref{fig:dip}, 
we see that the dip regions 
are not well fitted with a model in which 
the threshold cusps at the dips are removed from the default setting [blue dashed];
adding more $K^*_J$ in Table~\ref{tab:KJ} does not help. 
On the other hand,
the LHCb did not 
introduce resonances but use 
complicated interferences
to fit the dip regions.
Possibly due to 
this fitting choice,
the LHCb amplitude model actually needs significantly more mechanisms and fitting
parameters than our model does, as will be discussed shortly.

Another noteworthy point is that the LHCb's model includes a contact 
$B^+\to J/\psi \phi(1^+) K^+$ mechanism with a large ($\sim$28\%) 
fit fraction
while our model does not.
Since sequential two-body decay chains usually dominate, 
this large fit fraction could hint 
relevant missing mechanisms.
%
Although our model also includes contact mechanisms such as 
$B^+\to D_{sJ}^{(*)}\bar{D}_s^{(*)} K^+, D_{s}^{(*)}\bar{D}^{*} \phi$
in Fig.~\ref{fig:diag}(a-c),
they can be understood as color-favored 
sequential two-body decay chains such as 
$B^+\to D_{s(J)}^{(*)}\bar{D}^{\prime}$ followed by 
$\bar{D}^{\prime}\to \bar{D}_s^{(*)} K^+, \bar{D}^{*} \phi$,
and the off-shell excited charmed mesons ($\bar{D}^{\prime}$) in the
loops can be shrunk to the contact mechanisms.

We also point out the difference in the number of fitting parameters
($N_{\rm p}$) and its implication.
Our default model is fitted to 
the $M_{J/\psi\phi}$, $M_{J/\psi K^+}$, 
and $M_{K^+\phi}$ distributions with 
$N_{\rm p}=29$.
The LHCb's amplitude model is fitted to the six-dimensional distribution with
$N_{\rm p}=144$ and, in comparison with 
the $M_{J/\psi\phi}$, $M_{J/\psi K^+}$, and $M_{K^+\phi}$ distributions,
$\chi^2=82.5, 79.4, 60.7$, respectively.
This large difference in $N_{\rm p}$ should be partly from the fact that
the six-dimensional distribution include more information,
and that the LHCb's fit quality is somewhat better. 
However, this might not fully explain the difference in $N_{\rm p}$.
Possibly,
the LHCb's model misses 
relevant mechanisms
and needs many others to mimic the missing ones through
complicated interferences, resulting in the large $N_{\rm p}$.
At present,
we cannot discuss which of the LHCb's model or ours
is statistically more significant, 
since they were fitted to the different datasets.
%

Since the LHCb claimed $X(4630)(1^-)$ and $X(4150)(2^-)$, 
we added them to our default model to see their relevance. 
Although the fit quality is slightly improved
($\chi^2/{\rm ndf}=1.74$)
a similar improvement can also be made by $K^*_J$-excitation mechanisms
not included in the default model. 
We thus conclude that $X(4630)(1^-)$ and $X(4150)(2^-)$ are not relevant in
our model and their importance seems model-dependent, as far as we fit the
three invariant mass distributions.

Our default model fits well
the $Z_{cs}$-like structures 
in the $M_{J/\psi K^+}$ distribution
with the threshold cusps without any poles nearby. 
We examine 
to what extent a molecule (pole) scenario
for the $Z_{cs}$ structures is allowed 
by the LHCb data.
We vary the fitting parameters for
Fig.~\ref{fig:diag}(c) and also two independent 
$D_s^{+} \bar{D}^{*0}$ and 
$D_s^{*+} \bar{D}^{*0}$
 interaction strengths 
$h_\alpha$ in Eq.~(\ref{eq:cont-ptl}),
and find their allowed ranges.
For the $D_s^{+} \bar{D}^{*0}$ scattering, 
we find
$-0.33 < h_\alpha < 0.93$
that corresponds to the scattering length of 
$-0.12 < a {\rm (fm)} < 0.06$,
and a virtual pole at
93~MeV below the threshold or deeper.
Regarding the $D_s^{*+} \bar{D}^{*0}$ scattering, 
$-0.17 < h_\alpha < 2.02$, $-0.21 < a {\rm (fm)} < 0.03$,
and a virtual pole at 103~MeV below the threshold or deeper.

The result would disfavor the 
$D_s^{(*)+} \bar{D}^{*0}$ molecules as an explanation for the 
$Z_{cs}$ structures.
Meanwhile, Ortega et al.~\cite{zcs_model1} fitted well the 
$M_{J/\psi K^+}$ distribution with 
$D_s^{(*)+} \bar{D}^{*0}$
threshold cusps enhanced by 
virtual poles at $5-14$~MeV below the thresholds.
The difference from our result is partly from the fact that
they used momentum-independent
$D_s^{(*)+} \bar{D}^{*0}$ production vertices
while we used form factors.
If we also use momentum-independent production vertices,
we obtain, for the $D_s^{+} \bar{D}^{*0}$ scattering,
$-1.09 < h_\alpha < -0.25$, 
$0.04 < a {\rm (fm)} < 0.22$,
and a virtual pole at $48-99$~MeV below the threshold.
The molecule picture is still not clearly seen.
To further examine the molecule scenario, 
Ref.~\cite{fkguo15} stressed the importance of 
considering also
the elastic final state 
[e.g., the BESIII
$e^+ e^- \to K^+(D_s^- D^{*0}+D_s^{*-} D^0)$ data~\cite{BESIII:2020qkh}
in the present context].

The LQCD results~\cite{Prelovsek13,ychen14,Prelovsek15,ikeda16,cheung17}
suggested weak hadron-hadron interactions and neither bound nor
narrow resonances in the channel for $Z_c(3900)$ ($J^{PC}=1^{+-}$)
and its $1^{++}$ partner. 
Our results above, including the default model, are consistent with the 
LQCD results via the SU(3) relation;
most of the previous $Z_{cs}$ models did not take the consistency into
account. 
Yet, a non-pole scenario has not well explained
the experimentally observed
peak structures~\cite{BESIII:2020qkh,bes3_zx3900}
 that are commonly interpreted with
the $Z_c(3900)$ and  $Z_{cs}(3985)$ states.
More works from
experimental, phenomenological, and LQCD approaches 
are necessary to reach a consistent picture of 
$Z_{c(s)}$.

\begin{acknowledgments}
We thank F.-K. Guo for stimulating discussions and useful comments on
 the manuscript.
We also acknowledge L. Zhang for useful information on the LHCb
 amplitude analysis.
XL is supported by the National Natural
Science Foundation of China under Grants No. 12205002, and 
SXN is supported by 
National Natural Science Foundation of China (NSFC) under contracts 
U2032103 and 11625523, 
and also by
National Key Research and Development Program of China under Contracts 2020YFA0406400.
\end{acknowledgments}
\vspace{7mm}

\begin{center}
 {\bf Supplemental Material}
\end{center}

\begin{center}
 {1. Formulas for amplitudes in the default model}
\end{center}

We present one-loop amplitudes of Fig.~1(a)
that include
$s$-wave 
$D_{s0}^{*} \bar{D}_s(0^-)$, $D_{s1} \bar{D}_s^*(0^-)$, 
$D_{s0}^{*} \bar{D}_s^*(1^-)$ and $D_{s1} \bar{D}_s(1^-)$.
The initial weak vertices 
$B^+ \to D_{sJ}^{(*)} \bar{D}_s^{(*)}K^+$ are
\begin{align}
&c_{D_{s0}^* \bar{D}_s(0^-)} F^{00}_{D_{s0}^* \bar{D}_s K^+,B^+},
\\
&c_{D_{s1}\bar{D}_s^*(0^-)}\vec \varepsilon_{D_{s1}} \cdot \vec \varepsilon_{\bar{D}_s^*}F^{00}_{D_{s1} \bar{D}_s^* K^+,B^+},
\\
&c_{D_{s0}^* \bar{D}_s^*(1^-)} \vec p_{K^+} \cdot \vec \varepsilon_{\bar{D}_s^*} F^{01}_{D_{s0}^* \bar{D}_s^* K^+,B^+},
\\
&c_{D_{s1} \bar{D}_s(1^-)} \vec p_{K^+} \cdot \vec \varepsilon_{\bar{D}_{s1}} F^{01}_{D_{s1} \bar{D}_s K^+,B^+},
\end{align}
and the subsequent 
$D_{sJ}^{(*)} \bar{D}_s^{(*)} \to J/\psi \phi$ interactions are
\begin{align}
&c_{D_{s0}^*\bar{D}_s,\psi \phi}^{0^-} i\vec p_\phi \cdot (\vec \varepsilon_\psi \times \vec \varepsilon_\phi) f_{\psi\phi}^1 f_{D_{s0}^*\bar{D}_s}^0,
\\
&c_{D_{s1}\bar{D}_s^*,\psi\phi}^{0^-} \vec \varepsilon_{D_{s1}} \cdot \vec \varepsilon_{\bar{D}_s^*} i \vec p_\phi \cdot (\vec \varepsilon_\psi \times \vec \varepsilon_\phi) f_{\psi\phi}^1 f_{D_{s1}\bar{D}_s^*}^0,
\\
&c_{D_{s0}^*\bar{D}_s^*,\psi\phi}^{1^-} (\vec p_\phi \times \vec \varepsilon_{\bar{D}_s^*}) \cdot (\vec \varepsilon_\psi \times \vec \varepsilon_\phi)f_{\psi\phi}^1 f_{D_{s0}^*\bar{D}_s^*}^0,
\\
&c_{D_{s1}\bar{D}_s,\psi\phi}^{1^-} (\vec p_\phi \times \vec \varepsilon_{D_{s1}}) \cdot (\vec \varepsilon_\psi \times \vec \varepsilon_\phi)f_{\psi\phi}^1 f_{D_{s1}\bar{D}_s}^0,
\end{align}
respectively.
We have introduced dipole form factors $f_{ij}^{L}$, $f_{ij,k}^{L}$, and
$F_{ijk,l}^{LL'}$ given by
\begin{eqnarray}
\label{eq:ff2}
 f_{ij}^{L} &=&
 {1\over \sqrt{E_i E_j}}
\left(\frac{\Lambda^2}{\Lambda^2+q_{ij}^2}\right)^{2+(L/2)}, \\
f_{ij,k}^{L}&=& {f_{ij}^{L} \over \sqrt{E_k}}, \\
\label{eq:ff1}
 F_{ijk,l}^{LL'} &=&
 {f_{ij}^L\over \sqrt{E_k E_l}}
\left(\frac{\Lambda^{\prime 2}}{\Lambda^{\prime 2}+\tilde{p}_k^2}\right)^{\!\!2+{L'\over 2}}\!\!\!\!\!\!,
\end{eqnarray}
where 
$q_{ij}$ ($\tilde{p}_{k}$) 
is the momentum of $i$ ($k$) in the
$ij$ (total) center-of-mass frame;
$\Lambda^{(\prime)}$ is a cutoff for which
we use a common value of 1~GeV in all form factors.
With the above ingredients, the one-loop amplitudes are given by
\begin{widetext}
\begin{align}
&A_{D_{s0}^*\bar{D}_s(0^-)}^{\rm 1L}=c_{D_{s0}^*\bar{D}_s,\psi \phi}^{0^-} c_{D_{s0}^* \bar{D}_s(0^-)} i\vec p_\phi \cdot (\vec \varepsilon_\psi \times \vec \varepsilon_\phi) \int d^3 p_{\bar{D}_s} \frac{f_{\psi\phi}^1 f_{D_{s0}^*\bar{D}_s}^0 F^{00}_{D_{s0}^* \bar{D}_s K^+,B^+}}{M_{\psi\phi}-E_{D_{s0}^*}-E_{\bar{D}_s}+i\varepsilon},\label{eq:1}
\\
&A_{D_{s1} \bar{D}_s^*(0^-)}^{\rm 1L}=3c_{D_{s1} \bar{D}_s^*,\psi \phi}^{0^-}c_{D_{s1} \bar{D}_s^*(0^-)} i \vec p_\phi \cdot (\vec \varepsilon_\psi \times \vec \varepsilon_\phi) \int d^3 p_{\bar{D}_s^*} \frac{f_{\psi \phi}^1 f_{D_{s1} \bar{D}_s^*}^0 F_{D_{s1} \bar{D}_s^* K^+,B^+}^{00}}{M_{\psi \phi}-E_{D_{s1}}-E_{\bar{D}_s^*}+i\varepsilon},\label{eq:2}
\\
&A_{D_{s0}^* \bar{D}_s^*(1^-)}^{\rm 1L}=c_{D_{s0}^* \bar{D}_s^*,\psi \phi}^{1^-}c_{D_{s0}^* \bar{D}_s^*(1^-)} (\vec p_\phi \times \vec p_{K^+}) \cdot (\vec \varepsilon_\psi \times \vec \varepsilon_\phi) \int d^3 p_{\bar{D}_s^*} \frac{f_{\psi \phi}^1 f_{D_{s0}^* \bar{D}_s^*}^0 F_{D_{s0}^* \bar{D}_s^* K^+,B^+}^{01}}{M_{\psi \phi}-E_{D_{s0}^*}-E_{\bar{D}_s^*}+i\varepsilon},\label{eq:3}
\\
&A_{D_{s1} \bar{D}_s(1^-)}^{\rm 1L}=c_{D_{s1} \bar{D}_s,\psi \phi}^{1^-}c_{D_{s1} \bar{D}_s(1^-)} (\vec p_\phi \times \vec p_{K^+}) \cdot (\vec \varepsilon_\psi \times \vec \varepsilon_\phi) \int d^3 p_{\bar{D}_s} \frac{f_{\psi \phi}^1 f_{D_{s1} \bar{D}_s}^0 F_{D_{s1} \bar{D}_s K^+,B^+}^{01}}{M_{\psi \phi}-E_{D_{s1}}-E_{\bar{D}_s}+i\varepsilon}.\label{eq:4}
\end{align}
\end{widetext}

Similarly, one-loop amplitudes 
of Fig.~1(b) that include $s$-wave 
$D_s^{*} \bar{D}_s(1^+)$, $D_{s}^* \bar{D}_s^*(0^+)$, 
and $\psi^\prime \phi(0^+,1^+)$ are given by
\begin{widetext}
\begin{eqnarray}
\label{eq:01}
A^{\rm 1L}_{D_{s}^{*}\bar{D}_s(1^+)} &=&
c^{1^+}_{\psi\phi,D_s^*\bar{D}_s}\,
c_{D_s^*\bar{D}_s(1^+)}
i(\vec \varepsilon_\psi\times \vec \varepsilon_\phi)
\cdot \vec p_{K^+}
\int d^3p_{\bar{D}_s}
{ 
 f_{\psi\phi}^{0}
 f_{D_s^*\bar{D}_s}^{0}
 F_{D_s^*\bar{D}_s K^+,B}^{01}
\over M_{\psi \phi}-E_{D_s^*}-E_{\bar{D}_s}
+i\epsilon
}\ ,
\\
\label{eq:02}
A^{\rm 1L}_{D_{s}^{*}\bar{D}^*_s(0^+)} &=& 
 3\,c^{0^+}_{\psi\phi,D_s^*\bar{D}_s^*}\,
 c_{D_s^*\bar{D}_s^*(0^+)}\,
\vec \varepsilon_\psi\cdot \vec \varepsilon_\phi
\int d^3p_{\bar{D}^*_s}
{   
 f_{\psi\phi}^{0}
 f_{D_s^*\bar{D}^*_s}^{0} 
 F_{D_s^*\bar{D}_s^* K^+,B}^{00}
\over M_{\psi \phi}-E_{D_s^*}-E_{\bar{D}^*_s}
+i\epsilon
}\ ,
\\
\label{eq:03}
A^{\rm 1L}_{\psi'\phi(0^+)} &=& 
 3\,c^{0^+}_{\psi\phi,\psi'\phi}\,
 c_{\psi'\phi(0^+)}\,
\vec \varepsilon_\psi\cdot \vec \varepsilon_\phi
\int d^3p_{\psi'}
{   
 f_{\psi\phi}^{0}
 f_{\psi'\phi}^{0} 
 F_{\psi'\phi K^+,B}^{00}
\over M_{\psi \phi}-E_{\psi'}-E_{\phi}
+ {i\over 2}\Gamma_{\phi}
}\ ,
\\
\label{eq:04}
A^{\rm 1L}_{\psi'\phi(1^+)} &=& 
2\, c^{1^+}_{\psi\phi,\psi'\phi}\,
 c_{\psi'\phi(1^+)}\,
i (\vec \varepsilon_\psi\times \vec \varepsilon_\phi)\cdot\vec p_{K^+}
\int d^3p_{\psi'}
{   
 f_{\psi\phi}^{0}
 f_{\psi'\phi}^{0}
 F_{\psi'\phi K^+,B}^{01}
\over M_{\psi \phi}-E_{\psi'}-E_{\phi}
+ {i\over 2}\Gamma_{\phi}
}\ ,
\end{eqnarray}
\end{widetext}
respectively; $\Gamma_{\psi^\prime}$ has been neglected since
$\Gamma_{\psi^\prime} \ll \Gamma_\phi$.

Next we present 
one-loop amplitudes of 
Fig.~1(c) that include $s$-wave
$D_s\bar{D}^{*0} (1^+)$ and $D_s^*\bar{D}^{*0} (1^+)$. 
While 
parity-conserving 
and -violating weak decays can initiate the processes, 
our default model includes
parity-conserving 
$D_s\bar{D}^{*0} (1^+)$ 
and 
parity-violating
$D_s^*\bar{D}^{*0} (1^+)$
one-loop amplitudes.
The weak $B^+ \to D_s^{(*)}\bar{D}^{*0}\phi$ vertices
included in these amplitudes are
\begin{align}
&c_{D_s\bar{D}^{*0} (1^+)}^{\rm pc}\vec \varepsilon_{\bar{D}^{*0}} \cdot \vec \varepsilon_\phi F^{00}_{D_s\bar{D}^{*0} \phi,B},\label{eq:5}
\\
&c_{ D_s^*\bar{D}^{*0}(1^+)}^{\rm pv} (\vec \varepsilon_{\bar{D}^{*0}} \times \vec \varepsilon_{D_s^*}) \cdot (\vec p_\phi \times \vec \varepsilon_\phi) F^{01}_{D_s^*\bar{D}^{*0} \phi,B},\label{eq:6}
\end{align}
where superscripts pc and pv 
indicate parity-conserving and parity-violating, respectively.
The subsequent
$D_s^{(*)}\bar{D}^{*0}  \to J/\psi K^+$ interactions are given by
\begin{align}
&c_{K^+ \psi,D_s\bar{D}^{*0}}^{1^+} \vec \varepsilon_{\bar{D}^{*0}} \cdot \vec \varepsilon_{\psi} f_{K^+ \psi}^0 f^0_{D_s\bar{D}^{*0}},\label{eq:7}
\\ 
\nonumber \\
&c_{K^+ \psi,D_s^*\bar{D}^{*0}}^{1^+} i (\vec \varepsilon_{\bar{D}^{*0}} \times \vec \varepsilon_{D_s^*}) \cdot \vec \varepsilon_{\psi} f_{K^+ \psi}^0 f^0_{D_s^*\bar{D}^{*0}}.\label{eq:8}
\end{align}
With the above ingredients, the one-loop amplitudes are 
\begin{widetext}
\begin{align}
&A_{D_s\bar{D}^{*0}(1^+)}^{\rm 1L,pc}=c^{1^+}_{K^+\psi,D_s\bar{D}^{*0}}c_{D_s\bar{D}^{*0}(1^+)}^{\rm pc}\vec \varepsilon_\psi \cdot \vec \varepsilon_\phi \int d^3 p_{D_s} \frac{f^0_{K^+ \psi}f^0_{D_s\bar{D}^{*0}}F^{00}_{D_s\bar{D}^{*0}\phi,B}}{M_{\psi K^+}-E_{\bar{D}^{*0}}-E_{D_s}+i\varepsilon},\label{eq:9}
\\
&A_{D_s^*\bar{D}^{*0}(1^+)}^{\rm 1L,pv}=-2c^{1^+}_{K^+\psi,D_s^*\bar{D}^{*0}}c^{\rm pv}_{D_s^*\bar{D}^{*0}(1^+)}i(\vec \varepsilon_\psi \times \vec \varepsilon_\phi) \cdot \vec p_\phi \int d^3 p_{D_s^*} \frac{f^0_{K^+ \psi}f^0_{D_s^*\bar{D}^{*0}}F^{01}_{D_s^*\bar{D}^{*0}\phi,B}}{M_{\psi K^+}-E_{\bar{D}^{*0}}-E_{D_s^*}+i\varepsilon}.\label{eq:10}
\end{align}
\end{widetext}

Now we present formulas for $K^*_J$-excitation mechanisms of
Fig.~1(d); see Table~1 for $K^*_J$ considered in
our default model.
For $1^- K^*$,
our default model includes
a parity-conserving amplitude given by
\begin{align}\label{amp:1680}
A_{K^*}^{\rm pc}&=c_{K^*}^{\rm pc}\frac{(\vec p_\psi \times \vec \varepsilon_\psi) \cdot (\vec p_{\phi} \times \vec \varepsilon_\phi)f_{\phi K^+,K^*}^1 f_{K^* \psi,B}^1}{E-E_\psi-E_{K^*}+\frac{i}{2}\Gamma_{K^*}},
\end{align}
where $K^*$ is either $K^*(1410)$ or $K^*(1680)$.

As for $1^+ K_1$,
we consider both parity-conserving and -violating amplitudes given as
\begin{align}\label{amp:1650}
&A_{K_1}^{\rm pc}=c_{K_1}^{\rm pc}\frac{\vec \varepsilon_\psi \cdot \vec \varepsilon_\phi f_{\phi K^+,K_1}^0 f_{K_1\psi,B}^0}{E-E_\psi-E_{K_1}+\frac{i}{2}\Gamma_{K_1}},
\\ \label{amp:1650_1}
&A_{K_1}^{\rm pv}=c_{K_1}^{\rm pv}\frac{i(\vec \varepsilon_\psi \times \vec \varepsilon_\phi) \cdot \vec p_\psi f_{\phi K^+,K_1}^0 f_{K_1\psi,B}^1}{E-E_\psi-E_{K_1}+\frac{i}{2}\Gamma_{K_1}},
\end{align}
respectively, with $K_1=K_1(1650)$.

Regarding $2^- K_2$, 
our default model includes a parity-conserving amplitude given by
\begin{eqnarray}
\label{amp:1770}
A^{\rm pc}_{K_2}&=&C^{\rm pc}_{K_2}\frac{f_{\phi K^+,K_2}^1 f_{\psi K_2,B^+}^1}
{E-E_\psi-E_{K_2}+\frac{i}{2}\Gamma_{K_2}} 
 \left( \frac{1}{2}\vec \varepsilon_\phi \cdot \vec p_\psi \vec
 \varepsilon_\psi \cdot \vec p_\phi \right.
\nonumber\\
&&
\left.
+ \frac{1}{2}\vec \varepsilon_\phi \cdot \vec \varepsilon_\psi \vec
p_\phi \cdot \vec p_\psi 
-\frac{1}{3}\vec \varepsilon_\phi \cdot \vec p_\phi \vec
\varepsilon_\psi \cdot \vec p_\psi 
\right) ,
\end{eqnarray}
where $K_2$ is either $K_2(1770)$ or $K_2(1820)$.

In practice, we calculate amplitudes of Eqs.~(\ref{eq:1})-(\ref{eq:04})
[Eqs.~(\ref{eq:9}) and (\ref{eq:10})]
in the $J/\psi \phi$ [$J/\psi K^+$] center-of-mass frame.
The $K^*_J$-excitation amplitudes of Eqs.~(\ref{amp:1680})-(\ref{amp:1770})
are calculated in the total center-of-mass frame, but the second 
$K^{*+}_J\to K^+\phi$ vertices are calculated in the 
$K^+\phi$ center-of-mass frame, as in the helicity formalism
employed by the LHCb analysis. 
The invariant amplitudes are obtained 
from the above-presented amplitudes by multiplying 
relevant kinematical factors, and are plugged into the Dalitz plot
distribution formula;
see Appendix~B of Ref.~[52] for details.

Parameter values obtained from the fit
are listed in Table \ref{tab:para}. 
The masses and widths appearing in the above formulas are taken from
Ref.~[4].
In Table \ref{tab:para}, 
we also list each mechanism’s 
fit fraction defined by
\begin{align}\label{eq:fraction}
{\rm FF}.=\frac{\Gamma_{A_x}}{\Gamma_{\rm full}} \times 100 (\%),
\end{align}
where $\Gamma_{\rm full}$ and $\Gamma_{A_x}$ 
are $B^+ \to J/\psi \phi K^+$ decay rates 
calculated with the default model and 
with an amplitude $A_x$ only, respectively.
\begin{table*}[htbp]
\renewcommand{\arraystretch}{1.8}
\tabcolsep=3.mm
\caption{\label{tab:para}
Parameter values and fit fractions
for the default model ($\Lambda=1$~GeV)
obtained from fitting the LHCb data~[15].
The first column lists each
mechanism considered in our model, 
and the second column is 
its fit fraction (\%) defined in Eq.\eqref{eq:fraction}.
The third column lists
the product of coupling constants to fit the data, 
and its value and unit 
are given in the fourth and fifth columns, respectively.
Amplitude formulas are given in the equations in the last column.
}
\begin{tabular}[b]{lrlccc}
$A_{D_{s0}^*\bar{D}_s(0^-)}^{\rm 1L}$ &33.3 &$c_{D_{s0}^*\bar{D}_s,\psi \phi}^{0^-} c_{D_{s0}^* \bar{D}_s(0^-)}$ &$  -15.0 + 14.4 \,i$ &GeV$^{-1}$ &Eq.~(\ref{eq:1})  
\\
$A_{D_{s1} \bar{D}_s^*(0^-)}^{\rm 1L}$ &8.0 &$c_{D_{s1} \bar{D}_s^*,\psi \phi}^{0^-}c_{D_{s1} \bar{D}_s^*(0^-)}$ &$ 0.06+4.32 \,i$ &GeV$^{-1}$ &Eq.~(\ref{eq:2})   
\\
$A_{D_{s0}^* \bar{D}_s^*(1^-)}^{\rm 1L}$ &5.6 &$c_{D_{s0}^* \bar{D}_s^*,\psi \phi}^{1^-}c_{D_{s0}^* \bar{D}_s^*(1^-)}$  &$ -14.0-21.4 \,i$&GeV$^{-2}$ &Eq.~(\ref{eq:3})   
\\
$A_{D_{s1} \bar{D}_s(1^-)}^{\rm 1L}$ &5.2 & $c_{D_{s1} \bar{D}_s,\psi \phi}^{1^-}c_{D_{s1} \bar{D}_s(1^-)}$    &$ 21.6+10.8 \,i$&GeV$^{-2}$ &Eq.~(\ref{eq:4})  
\\
$A_{D_s^*\bar{D}_s(1^+)}^{\rm 1L}$  &16.5& $c^{1^+}_{\psi\phi,D_s^* \bar{D}_s} c_{D_s^*\bar{D}_s(1^+)}$    &$ 12.7+2.26 \,i$ &GeV$^{-1}$ &Eq.~(\ref{eq:01})  
\\
$A_{D_s^*\bar{D}_s^*(0^+)}^{\rm 1L}$ &7.4 &$c^{0^+}_{\psi\phi,D_s^* \bar{D}_s^*} c_{D_s^*\bar{D}_s^*(0^+)}$    &$  0.08+1.24 \,i$&--- &Eq.~(\ref{eq:02})  
\\
$A_{\psi^\prime \phi(0^+)}^{\rm 1L}$ &15.1 &$c^{0^+}_{\psi\phi,\psi^\prime \phi} c_{\psi^\prime \phi(0^+)}$   &$ 4.04$&--- &Eq.~(\ref{eq:03})  
\\
$A_{\psi^\prime \phi(1^+)}^{\rm 1L}$ &14.4 &$c^{1^+}_{\psi\phi,\psi^\prime \phi} c_{\psi^\prime \phi(1^+)}$ &$  16.6	-2.14\,i$ &GeV$^{-1}$ &Eq.~(\ref{eq:04}) 
\\
$A_{D_s\bar{D}^{*0}(1^+)}^{\rm 1L,pc}$ &9.8 &$c^{1^+}_{K^+\psi,D_s\bar{D}^{*0}}c_{D_s\bar{D}^{*0}(1^+)}^{\rm pc}$ &$ -3.34-3.56  \,i$&--- &Eq.~(\ref{eq:9})   
\\
$A_{D_s^*\bar{D}^{*0}(1^+)}^{\rm 1L,pv}$ &2.4 &$c^{1^+}_{K^+\psi,D_s^*\bar{D}^{*0}}c^{\rm pv}_{D_s^*\bar{D}^{*0}(1^+)}$ & $  2.97 \,i$&GeV$^{-1}$ & Eq.~(\ref{eq:10})
\\
$A_{K^*(1680)}^{\rm pc}$ &37.5 &$c_{K^*(1680)}^{\rm pc}$  &$   -28.4+38.0 \,i$ &--- & Eq.~(\ref{amp:1680})\\
$A_{K^*(1410)}^{\rm pc}$ &49.1 &$c_{K^*(1410)}^{\rm pc}$  &$-25.4-62.6 \,i$ &--- &  Eq.~(\ref{amp:1680})\\
$A_{K_1(1650)}^{\rm pc}$ &10.9 &$c_{K_1(1650)}^{\rm pc}$ &$   0.97+2.41 \,i$ &GeV$^{2}$ & Eq.~(\ref{amp:1650}) \\
$A_{K_1(1650)}^{\rm pv}$ &18.8 &$c_{K_1(1650)}^{\rm pv}$ &$   5.93+1.06 \,i$ &GeV$^{1}$ & Eq.~(\ref{amp:1650_1}) \\
$A^{\rm pc}_{K_2(1770)}$ &9.0 &$c_{K_2(1770)}^{\rm pc}$&$   -24.3+6.03 \,i$ &--- & Eq.~(\ref{amp:1770}) \\
$A^{\rm pc}_{K_2(1820)}$ &20.0 &$c_{K_2(1820)}^{\rm pc}$ &$   31.9+23.4 \,i$ &--- & Eq.~(\ref{amp:1770})
\end{tabular}
\end{table*}

\end{document}